\documentclass[aip, amsmath,amssymb, reprint, floatfix]{revtex4-1}
\usepackage{dcolumn}
\usepackage{color}
\usepackage{graphicx}
\usepackage{bm}
\usepackage{soul}
\usepackage[utf8]{inputenc}
\usepackage[T1]{fontenc}
\usepackage{mathptmx}
\usepackage{natbib}
\usepackage{multirow}
\usepackage{colortbl}
\usepackage{chngcntr}
\usepackage{upgreek}

\definecolor{BlackAuto}{rgb}{0,0,0}

\definecolor{Grayy}{rgb}{0.9,0.9,0.9}
\definecolor{Mauve}{rgb}{1,0.9,1}
\definecolor{ligthgreen}{rgb}{0.9,1,0.4}
\definecolor{OliveGreen}{rgb}{0,0.6,0}
\definecolor{NavyBlue}{rgb}{0.0, 0.07, 0.8}
\definecolor{ModeBeige}{rgb}{0.59, 0.44, 0.09}
\definecolor{MordantRed}{rgb}{1, 0.0, 0.0}
\definecolor{Orange}{rgb}{1.0, 0.5, 0.0}
\definecolor{SkyMagenta}{rgb}{0.81, 0.44, 0.69}
\definecolor{BlueGreen}{rgb}{0.0, 0.55, 0.5}
\definecolor{BlueViolet}{rgb}{0.54, 0.17, 0.89}
\definecolor{Indigo}{rgb}{0.47, 0.11, 0.97}
\definecolor{Caramel}{rgb}{0.49, 0.2, 0}

\usepackage{amsmath}
\usepackage{amsfonts}
\usepackage{amssymb}

\newcommand{\titre}{Control of the magnetic anisotropy in multi-repeat Pt/Co/Al heterostructures using magneto-ionic gating}

\begin{document}

\title{\titre}

\author{Tristan da Câmara Santa Clara Gomes}
\email{tristan.dacamara@uclouvain.be. Present address: Institute of Condensed Matter and Nanosciences, Universit\'{e} catholique de Louvain, Place Croix du Sud 1, 1348 Louvain-la-Neuve, Belgium.}
\affiliation{Laboratoire Albert Fert, CNRS, Thales, Universit\'e Paris-Saclay, 91767 Palaiseau, France}
\author{Tanvi Bhatnagar-Schöffmann}
\affiliation{Centre de Nanosciences et de Nanotechnologies, CNRS, Universit\'e Paris-Saclay, 91120 Palaiseau, France}
\author{Sachin Krishnia}
\affiliation{Laboratoire Albert Fert, CNRS, Thales, Universit\'e Paris-Saclay, 91767 Palaiseau, France}
\author{Yanis Sassi}
\affiliation{Laboratoire Albert Fert, CNRS, Thales, Universit\'e Paris-Saclay, 91767 Palaiseau, France}
\author{Dedalo Sanz-Hern\'{a}ndez}
\affiliation{Laboratoire Albert Fert, CNRS, Thales, Universit\'e Paris-Saclay, 91767 Palaiseau, France}
\author{Nicolas Reyren}
\affiliation{Laboratoire Albert Fert, CNRS, Thales, Universit\'e Paris-Saclay, 91767 Palaiseau, France}
\author{Marie-Blandine Martin}
\affiliation{Laboratoire Albert Fert, CNRS, Thales, Universit\'e Paris-Saclay, 91767 Palaiseau, France}
\author{Frederic Brunnett}
\affiliation{Laboratoire Albert Fert, CNRS, Thales, Universit\'e Paris-Saclay, 91767 Palaiseau, France}
\author{Sophie Collin}
\affiliation{Laboratoire Albert Fert, CNRS, Thales, Universit\'e Paris-Saclay, 91767 Palaiseau, France}
\author{Florian Godel}
\affiliation{Laboratoire Albert Fert, CNRS, Thales, Universit\'e Paris-Saclay, 91767 Palaiseau, France} 
\author{Shimpei Ono}
\affiliation{Central Research Institute of Electric Power Industry, Yokosuka, Kanagawa 240-0196, Japan}
\author{Damien Querlioz}
\affiliation{Centre de Nanosciences et de Nanotechnologies, CNRS, Universit\'e Paris-Saclay, 91120 Palaiseau, France}
\author{Dafiné Ravelosona}
\affiliation{Centre de Nanosciences et de Nanotechnologies, CNRS, Universit\'e Paris-Saclay, 91120 Palaiseau, France}
\author{Vincent Cros}
\affiliation{Laboratoire Albert Fert, CNRS, Thales, Universit\'e Paris-Saclay, 91767 Palaiseau, France}
\author{Julie Grollier}
\affiliation{Laboratoire Albert Fert, CNRS, Thales, Universit\'e Paris-Saclay, 91767 Palaiseau, France}
\author{Pierre Seneor}
\affiliation{Laboratoire Albert Fert, CNRS, Thales, Universit\'e Paris-Saclay, 91767 Palaiseau, France}
\author{Liza Herrera Diez}
\affiliation{Centre de Nanosciences et de Nanotechnologies, CNRS, Universit\'e Paris-Saclay, 91120 Palaiseau, France}

\begin{abstract}
Controlling magnetic properties through the application of an electric field is a significant challenge in modern nanomagnetism. In this study, we investigate the magneto-ionic control of magnetic anisotropy in the topmost Co layer in Ta/Pt/[Co/Al/Pt]$_n$/Co/Al/AlO$_\text{x}$ multilayer stacks comprising $n +1$ Co layers and its impact on the magnetic properties of the multilayers. We demonstrate that the perpendicular magnetic anisotropy can be reversibly quenched through gate-driven oxidation of the intermediary Al layer between Co and AlO$_\text{x}$, enabling dynamic control of the magnetic layers contributing to the out-of-plane remanence — varying between $n$ and $n +1$. For multilayer configurations with $n = 2$ and $n = 4$, we observe reversible and non-volatile additions of 1/3 and 1/5, respectively, to the anomalous Hall effect amplitude based on the applied gate voltage. Magnetic imaging reveals that the gate-induced spin-reorientation transition occurs through the propagation of a single 90$^{\circ}$ magnetic domain wall separating the perpendicular and in-plane anisotropy states. In the 5-repetition multilayer, the modification leads to a doubling of the period of the magnetic domains at remanence. These results demonstrate that the magneto-ionic control of the anisotropy of a single magnetic layer can be used to control the magnetic properties of coupled multilayer systems, extending beyond the gating effects on a single magnetic layer. 
\end{abstract}

\maketitle

Electric field control of magnetism has emerged as a promising avenue for manipulating magnetic states without the need for traditional current-induced methods. This novel opportunity shall allow to develop energy-efficient data storage and novel functionalities for neuromorphic computing devices \cite{Weisheit2007, Bernand-Mantel2013, Matsukura2015, Song2017, Schott2017, Zahedinejad2022}. Of the various approaches for electric-field control of magnetism, magneto-ionic control, based on the motion of ions induced by the electric field, has garnered significant attention due to its potential for non-volatile tuning of magnetic properties \cite{Bernand-Mantel2013, Bi2014, Liu2017, Srivastava2018, HerreraDiez2019, Fassatoui2020}. Magneto-ionic gating of key magnetic properties such as the magnetic anisotropy or the interfacial Dzyaloshinskii-Moriya interaction (DMI), allows for the controlled manipulation of static domain wall patterns or dynamic domain wall nucleation and motion \cite{Bernand-Mantel2013, Liu2017, Schott2017, Srivastava2018, Guan2021, Balan2022, Fillion2022, Dai2023}. Extensive research has been conducted to enhance the stability and reversibility of non-volatile states and gain insights into the influence of ionic motion on magnetic properties, notably, optimizing the materials for the ionic gates as well as the composition and interfaces of magnetic layers of high technological relevance \cite{Fassatoui2020, Lee2020, Tanvi2023, Pachat2022}. In particular, oxygen-based magneto-ionics, in which the gate-induced modulation of the content of oxygen species near the interface with the magnetic layer can be used to fine-tune perpendicular magnetic anisotropy (PMA), is one of the most developed methods for implementing magneto-ionic control of interfacial magnetic properties in nanostructures \cite{Bauer2015, Pachat2021, Fillion2022}.

Magneto-ionics traditionally modifies interfacial properties in thin films with a single magnetic layer. However, multilayer structures offer unique advantages for device functionality. By strategically layering repetitive sequences of magnetic and non-magnetic layers, both the magnetic anisotropy and DMI can be effectively fine-tuned. Typically, these structures alternate between magnetic and heavy metal and/or oxide layers, capitalizing on both strong interface contributions and the combined effects of successive interfaces. This stack design has notably been employed to stabilize columnar skyrmions in Co layers, that remain stable at room temperature, even for sub-100\,nm skyrmion diameter \cite{Fert2017, Moreau-Luchaire2016, Woo2016, Yu2016, Soumyanarayanan2017, Legrand2017}. This strategy offers therefore a promising pathway to design materials where domain wall and skyrmion nucleation, motion, and annihilation, can be electrically controlled, representing a crucial step toward realizing advanced spintronics devices 
\cite{Fert2017, Legrand2017, Hrabec2017, TristanUnpublished}. Notably, magnetic multilayers composed of several thin repetitions of Pt(3\,nm)/Co(1.2\,nm)/Al(3\,nm) provide suitable magnetic anisotropy and DMI to stabilize magnetic skyrmions \cite{Ajejas2022}, while also offering strong spin-orbit torque \cite{krishnia2022large} from both the Pt/Co interface and the Co/Al interface, leading to efficient skyrmion motion. Therefore, implementing a pathway towards the magneto-ionic control in such complex multilayered structures presents an innovative opportunity for non-volatile electrical control of key magnetic properties going beyond the gating effects induced on a single layer.

In this work, we investigate the magneto-ionic modulation of effective magnetic anisotropy in multi-repeat sputtered Pt/Co/Al layers. We demonstrate a reversible, non-volatile transition of the magnetic anisotropy of the topmost Co layer from in-plane to out-of-plane using a magneto-ionic gate voltage. For repetitions with $n =$ 2 and $n =$ 4, this enables control over the total number of active PMA magnetic layers, denoted as $m$, shifting between $m = n$ and $m = n +$1 layers. We also show that the gate-voltage induced spin-reorientation transition between PMA and in-plane magnetic anisotropy occurs through the propagation of a single magnetic domain wall (DW$_{\updownarrow,\leftrightarrow}$) front, driven by the applied electric field. We observe that this DW$_{\updownarrow,\leftrightarrow}$ does not interact significantly with the magnetic multidomain configuration of the underlaying coupled Pt/Co/Al layers with PMA. 

\begin{figure}
    \centering
    \includegraphics[scale=0.69]{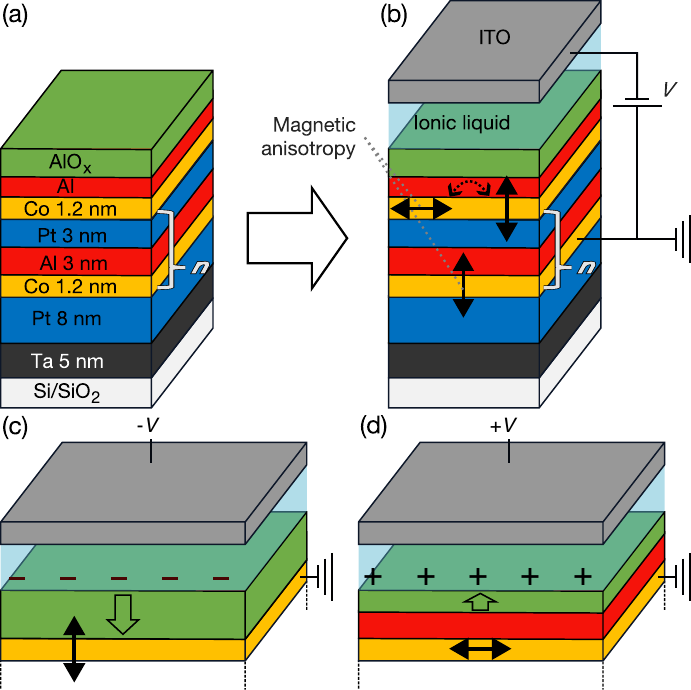}
    \caption{(a-b) Schematic of the (a) Ta/Pt/[Co/Al/Pt]$_n$/Co/Al/AlO$_\text{x}$ multilayer stacks and (b) the magneto-ionic devices allowing for the control of the magnetic anisotropy in the topmost Co layer. \textcolor{black}{(c-d) Schematics of voltage gate control of the oxidation front in the Al/AlO$_\text{x}$ capping layers, which can be displaced downward or upward using negative (c) or positive (d) gate voltages applied to the top ITO electrode.}}
    \label{FigureIntro}
\end{figure}

The magnetic stacks used in this study \textcolor{black}{are illustrated in Fig.~\ref{FigureIntro} (a). They }consist of $n$ repetitions of [Co(1.2\,nm)/Al(3\,nm)/Pt(3\,nm)], which are deposited on top of a Ta(5\,nm)/Pt(8\,nm) buffer layer. The topmost layers consist of Co(1.2\,nm)/Al($y$\,nm) where \textcolor{black}{$y$ refers to the deposited Al layer thickness, which has been varied between 0.6 to 1\,nm. Subsequently, the top Al layer is oxidized in air, forming a top Al/AlO$_\text{x}$ bilayer, as depicted in Fig.~\ref{FigureIntro} (a).} All stacks are grown at room temperature by magnetron sputtering on thermally oxidized silicon substrate (280\,nm of SiO$_2$). Ionic liquid gating is achieved using the ionic liquid 1-Ethyl-3-methylimidazolium bis(trifluoromethanesulfonyl)imide [EMI]$^+$ [TFSI]$^-$ dropped on the surface of the magnetic stacks \cite{Ono2008}. A glass coated with Indium Tin Oxide (ITO) is placed on top of the ionic liquid to serve as counter electrode, as illustrated in Fig.~\ref{FigureIntro} (b). All indicated gate voltages correspond to the voltages applied to the top ITO electrode, with the magnetic layers grounded. The exposed area to the gate voltage is approximately 0.25 cm$^2$. The non-volatile magnetic anisotropy changes induced by magneto-ionic gating have been measured through Anomalous Hall Effect (AHE) after the gate voltage is switched off. Alternative Gradient Field Magnetometer (AGFM) measurements have been performed after completely removing the ionic liquid - ITO gate and, to assess the non-volatile character, up to several weeks after the application of the gate voltage. Kerr microscopy measurements have been performed during and after the application of electric field, as well as after the removal of the gate. All measurements have been conducted at room temperature.
 
\begin{figure*}
\centering
\includegraphics[scale=0.7]{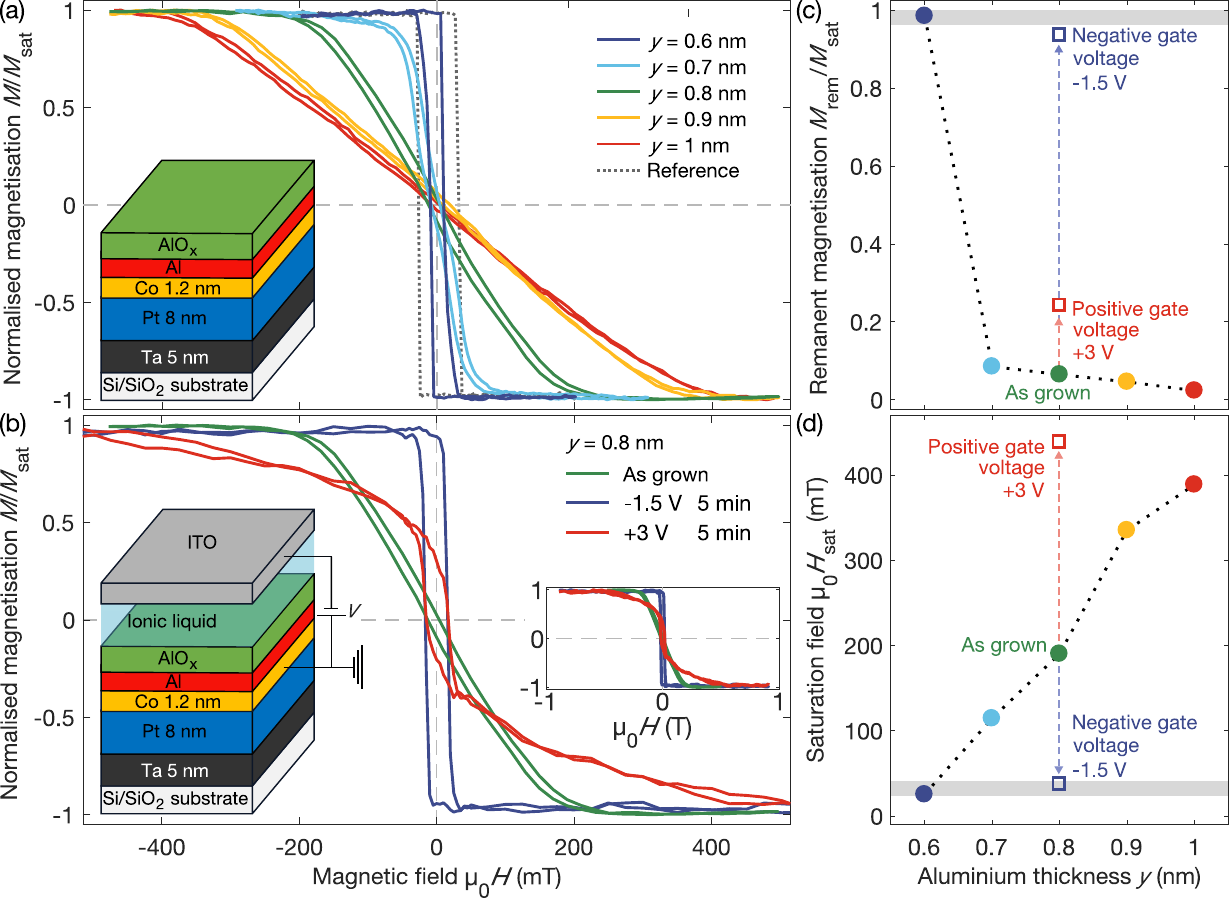}
\caption{(a) Out-of-plane hysteresis curves obtained by Kerr microscopy for Ta/Pt/Co/Al/AlO$_\text{x}$ multilayer stacks, where \textcolor{black}{the nominally deposited Al capping layer thickness} $y$ ranges from 0.6 to 1\,nm, inducing significant changes in the magnetic anisotropy. (b) Normalized Kerr signal for the Ta/Pt/Co/Al(0.8\,nm) stack in the as-grown state, and after the application of gate voltages of $-$1.5\,V for 5\,minutes and subsequently $+$3\,V, for 5\,minutes showing that gate voltages produce an equivalent effect in anisotropy as a variation in $y$. The inset in (b) show the same data at higher magnetic fields, providing a better view of the saturation field after positive gate voltage. (c-d) Normalized remanent Kerr signal (c) and saturation field $\mu_0H_\text{sat}$ (d) in the out-of-plane direction as a function of $y$. For the sample with $y =$ 0.8\,nm, the values after negative ($-$1.5\,V) and positive ($+$3\,V) gate voltage applications are also indicated. The results corresponding to a reference sample of Ta/Pt/Co/Al(3\,nm)/Pt are shown as a dotted curve in (a), and as grey areas in (c) and (d).}
\label{Fig1CoLayer}
\end{figure*}

In order to evaluate the impact of the oxygen content at the Co/Al interface on its magnetic anisotropy prior to the gate-voltage application, a series of single Ta(5\,nm)/Pt(8\,nm)/Co(1.2\,nm)/Al($y$\,nm) layers, in which $y$ represents the thickness of the top Al layer, are characterised. The variation in $y$ is introduced as a means to evaluate the effects of the proximity of the AlO$_\text{x}$ oxidation front, created by the exposure to air, to the Co/Al interface. In Figure~\ref{Fig1CoLayer} (a), we show the out-of-plane (OOP) hysteresis loops obtained by Kerr microscopy for a series of multilayer stacks with $y$ ranging from 0.6 to 1\,nm. We observe a strong dependence of the magnetic anisotropy on $y$. More precisely, PMA is observed for $y =$ 0.6\,nm, characterized by a distinct square-shaped loop exhibiting significant remanence in the OOP direction. For thicker top Al layers, PMA is reduced, as shown by the reduced squareness and remanent magnetisation of the corresponding curves. For $y =$ 0.8\,nm, the easy axis becomes the IP direction (see the AGFM measurements in Supplementary Information \cite{SuppMat} Fig.~\textbf{S2}). Then, the IP magnetic anisotropy keeps increasing for $y =$ 0.9 \,nm and 1\,nm. This behaviour is also reflected in the saturation field along the OOP direction $\mu_0H_\text{sat}$, which is found to increase almost linearly with $y$ from about 25\,mT for the $y =$ 0.6\,nm stack to 389\,mT for the $y =$ 1\,nm stack. These changes are summarized in Fig.~\ref{Fig1CoLayer} (c) and (d), showing the dependence of the remanent magnetization (c) and $H_\text{sat}$ (d) in the out-of-plane direction as a function of $y$. The strong variation of the magnetic anisotropy as a function of $y$ can be attributed to different positions of the oxidation front within the partially oxidized upper Al layer with respect to the Co surface, which modifies the anisotropy contribution from the top Co layer interface. It is widely documented in the literature, that the oxygen content at the Co/Al interface has a critical impact in magnetic anisotropy, since a strong interfacial perpendicular anisotropy is promoted by the hybridization between Co \textit{d} and O \textit{sp} orbitals \cite{Rodmacq2003, Lacour2007, Manchon2008, Dahmane2009}. However, only a narrow window of oxygen content produces PMA, under-oxidation as well as over-oxidation are detrimental to PMA, which contributes to the high sensitivity of this system to small variations of the oxygen content at the Co surface \cite{Rodmacq2003, Manchon2008}. \textcolor{black}{Our results align with this scenario, suggesting full oxidation of the Al layer when 0.6\,nm of Al is deposited and subsequently exposed to air. However, for thicker Al layers, air-induced oxidation results in an under-oxidized Al layer, leading to an in-plane magnetic anisotropy in the underlying Co layer. }Interestingly, reports in the literature also show that after this oxidation-dominated regime of spin reorientation transitions between PMA and IP anisotropies, PMA does not necessarily show an asymptotic decrease for higher Al thicknesses. At relatively high Al thicknesses, a spin orientation transition that is unrelated to the oxygen content at the Co interface can also occur mediated by complex interfacial electronic effects at the Co/metallic Al interface \cite{krishnia2022large, Ajejas2022}. In the present case, a similar scenario is observed, the monotonic decrease of PMA going from $y =$ 0.6\,nm to $y =$ 1\,nm, is followed by a recovery of PMA for $y =$ 3\,nm, labelled as reference. Figure~\ref{Fig1CoLayer} (a), (c) and (d) present data of a reference Ta(5\,nm)/Pt(8\,nm)/Co(1.2\,nm)/Al(3\,nm)/Pt(3\,nm) sample, which exhibits a $\mu_0H_\text{sat}$ value similar to that of the sample with $y =$ 0.6\,nm (without the Pt capping layer). The structure of the reference sample has been used as the single unit that is repeated $n$ times to form multilayer structures with PMA.

In Fig.~\ref{Fig1CoLayer} (b), we show the OOP Kerr effect loops obtained for the layer with $y =$ 0.8\,nm. In its as-grown state, this sample exhibits an IP magnetic anisotropy with $\mu_0H_\text{sat} =$ 190\,mT. After the application of a gate voltage of $-$1.5\,V for 5\,min, the PMA of the layer is increased, evidenced by a reduction of $\mu_0H_\text{sat}$ to 38\,mT, resembling the results obtained for $y =$ 0.6\,nm. A subsequent application of $+$3\,V for 5\,min to the same sample reduces PMA well below its value in the as-grown state, evidenced by a significant increase of $\mu_0H_\text{sat}$ to 439\,mT, closer to the results obtained for layers with $y =$ 1\,nm. The small-amplitude hysteresis and remanence observed in this case is attributed to a difference in the positioning of the gate between each loop \footnote{Each loop is taken after the removal of the ionic liquid gate to confirm the non-volatile modification of the anisotropy, therefore, the repositioning of the ionic liquid and ITO glass, as well as the signal contribution from the contact regions can lead to contributions from unbiased areas, as already shown in other ionic-liquid gate systems \cite{Pachat2022}. } These changes induced by the applied voltage are non-volatile and reversible. The reversibility during several successive cycles of positive and negative gate voltage application is shown in the supplementary information (see Supplementary Information \cite{SuppMat} Fig.~\textbf{S3}). These results are summarised in Fig.~\ref{Fig1CoLayer} (c) and (d) for $y =$ 0.8\,nm. \textcolor{black}{It should be noted that the curves presented in Fig.~\textbf{S3} exhibit no reduction in the saturation in the saturation of the AHE after the application of a negative gate voltage. This observation suggests that the gate driven changes in anisotropy happen without apparent changes in the oxidation state of the Co, and therefore rely on a significant contribution from the gate-driven motion of the oxidation front inside the Al layer.} 

The observed gate-driven modification of the anisotropy can be explained by the non-volatile gate-voltage-induced migration of oxygen species from and away from the Co/Al interface. This can be seen as a controlled displacement of the oxidation front located at the Al/AlO$_\text{x}$ interface towards the Co/Al interface for negative gate voltages applied to the top ITO electrode, and the opposite for positive gate voltages \textcolor{black}{(as schematically illustrated in Fig.~\ref{FigureIntro} (c-d)}. As shown in the results presented in Fig.~\ref{Fig1CoLayer}, this gate-driven process has effects on anisotropy that are equivalent to changing the thickness of the Al layer $y$, where higher PMA is obtained for thinner Al layers (thicker AlO$_\text{x}$ through the gate-induced migration of oxygen species towards the Co/Al interface), and vice versa. This equivalence between gating effects and the thickness of the metallic Al layer has already been reported for charge-based gating effects in Pt/Co/Al(wedge)/AlO$_\text{x}$/HfO$_\text{2}$ solid-state devices \cite{Bernand-Mantel2013}. However, unlike volatile charge-based effects, magneto-ionic effects can be directly associated with gate-induced diffusion of oxygen species\textcolor{black}{, leading to a resilient and non-volatile modification of magnetic anisotropy.}

\begin{figure}
\centering
\includegraphics[scale=0.7]{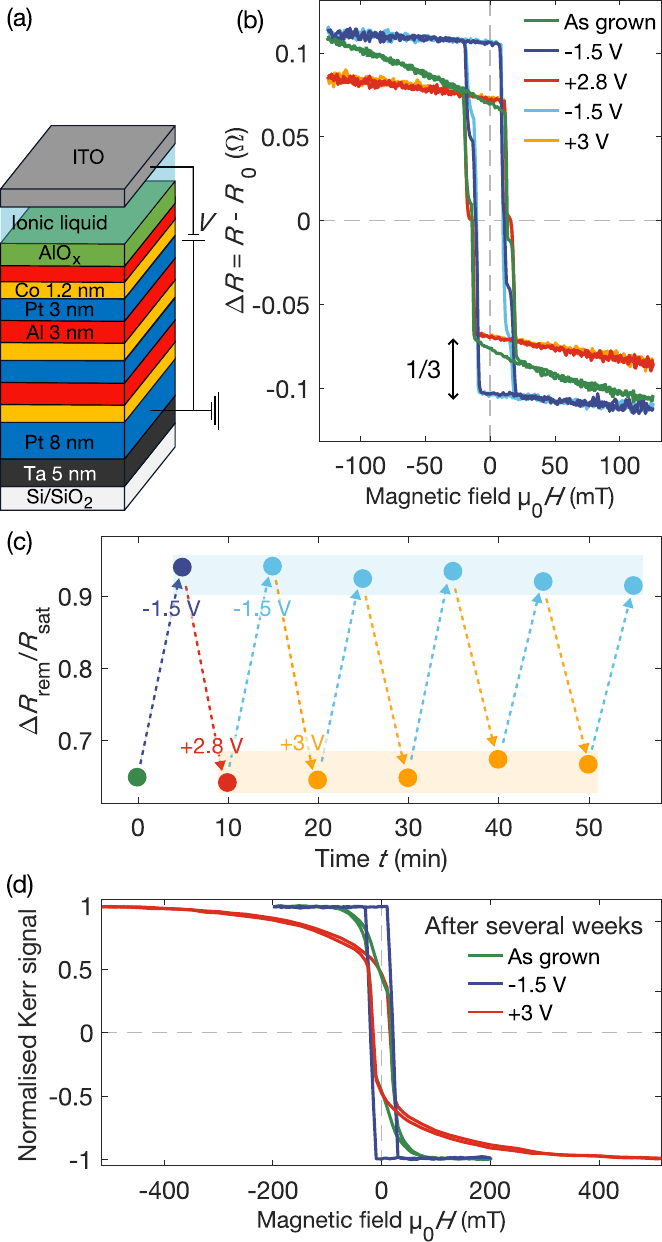}
\caption{(a) Magneto-ionic device composed of a Ta/Pt/[Co/Al/Pt]$_2$/Co/Al/AlO$_\text{x}$ ($y$ = 0.8\,nm) multilayer stack. (b) Anomalous Hall resistance $\Delta R = R  - R_0$, measured in the as-grown state and after the application of alternating negative and positive gate voltages. (c) Evolution of the normalised remanent Hall resistance $\Delta R_\text{rem}/R_\text{sat}$ with the gate voltage application time $t$ for positive and negative gate voltages. (d) OOP normalized magnetisation curves obtained by Kerr microscopy for the as-grown state, as well as after the application of positive and negative gate voltages. These later measurements are performed after the  removal of the gate.}
\label{Fig3CoLayer}
\end{figure}

After having studied the impact of magneto-ionic effects in single magnetic films, we consider the case of the multi-repeat stacks composed of three and five layers of Co ($n =$ 2 and $n =$ 4), in both cases the thickness of the topmost Al layer that is let to oxidize in air is $y$ = 0.8\,nm. In Figure~\ref{Fig3CoLayer} (a), we depict a schematic of a magneto-ionic device made of a Ta(5\,nm)/Pt(8\,nm)/[Co(1.2\,nm)/Al(3\,nm)/Pt(3\,nm)]$_2$
/Co(1.2\,nm)/\textcolor{black}{Al/AlO$_\text{x}$ ($y$ = 0.8\,nm)} stack. In Figure~\ref{Fig3CoLayer} (b), we show the anomalous Hall resistance loops shifted by the offset resistance value $R_0$, defined as $\Delta R = R - R_0$. The hysteresis loops are obtained using four electrical contacts on each corner of the stack, enclosing the gated area. In the as-grown state, the individual contribution of each of the three Co layers can be observed. The total anomalous Hall resistance amplitude $\Delta R$ can be subdivided in a first range varying monotonically with the external OOP magnetic field and two subsequent sections defined by abrupt transitions found for magnetic fields of about 13\,mT and 18\,mT, the $\Delta R$ amplitude of each of these three contributions accounting for about one third of the total signal. The range showing a monotonic variation with the external OOP magnetic field is assigned to the top Co layer in contact to the oxidized Al(0.8\,nm) layer. As shown in Fig.~\ref{Fig1CoLayer} (b), this layer does not have a strong PMA, therefore the observed quasi-linear variation of $\Delta R$ with the external OOP magnetic field can be attributed to a coherent spin rotation towards the OOP hard-axis. In contrast, the two bottom Co reference layers are expected to show strong PMA, as seen in Fig.~\ref{Fig1CoLayer} (a), which is confirmed by the observation of the sharp switching of $\Delta R$. The small differences of switching field can be attributed to the difference of the thickness of the Pt layers on which Co layers are grown. The bottom Co layer is grown on top of Pt(8\,nm), while the middle layer is grown on top of Pt(3\,nm). This is confirmed by the results obtained for Ta(5\,nm)/Pt(8\,nm)/[Al(3\,nm)/Pt(3\,nm)/Co(1.2\,nm))]$_3$/Al(0.8 nm) multilayer stacks. In this case, all Co layers are grown on a 3\,nm thick Pt layer and only one abrupt switching accounting for about two third of the total $\Delta R$ variation is observed (see Supplementary Information \cite{SuppMat} Fig.~\textbf{S4}).

\begin{figure*}
\centering
\includegraphics[scale=0.74]{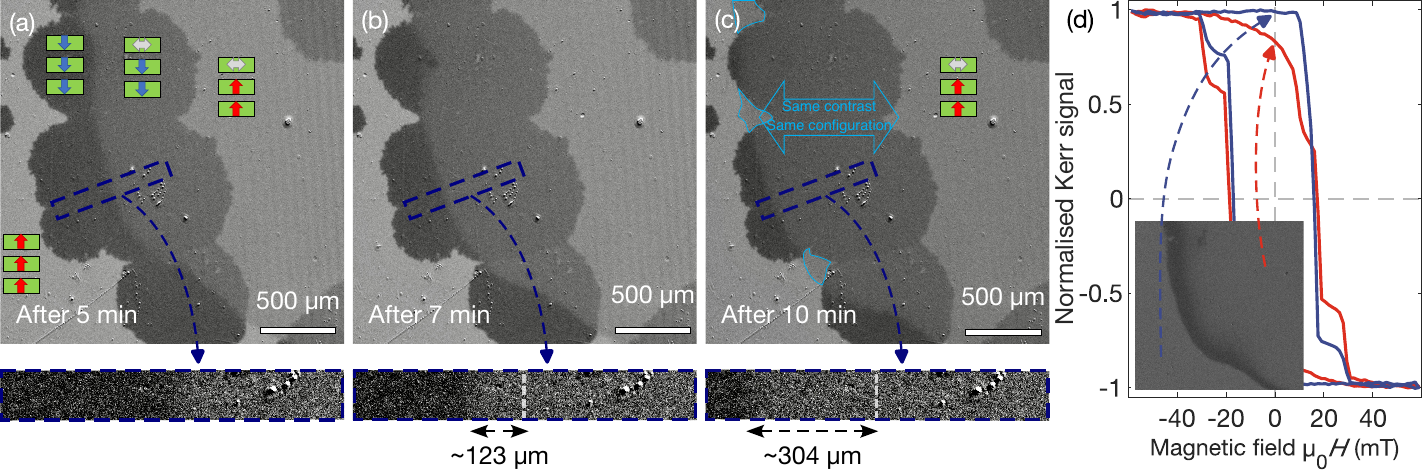}
\caption{(a-c) Kerr microscopy images of the domain configuration during the application of positive gate voltage of $+$2.5\,V during 5\,min (a), 7\,min (b) and 10\,min (c), showing the transition behavior of the Ta/Pt/[Co/Al/Pt]$_2$/Co/Al/AlO$_\text{x}$ ($y$ = 0.8\,nm) multilayer stack. The bottom insets show zoom at the domain wall between PMA and IP magnetic anisotropy domains in the topmost Co layer moving at a velocity of 61\,$\upmu$m/min. \textcolor{black}{(a) The schematic magnetization configuration of the three Co layers is displayed over each different Kerr contrast. (c) Due to the progression of the IP domain compared to the fixed OOP ones, some of the contrast observed in the right-most part of the image appears on the other side of the large OOP domain with magnetization pointing down in the middle of the image.} (d) Hysteresis loops measured at the top-down corner (blue), and over the entire field of view (red) of the image in the inset. The inset in (d) shows an extended view of the magnetic state in (c) at remanence.}
\label{Fig3CoDW}
\end{figure*}

As shown already in Fig.~\ref{Fig1CoLayer} (b), the application of a gate voltage of $-$1.5\,V during 5\,min induces a spin-reorientation transition of the magnetization of the topmost Co layer towards the perpendicular direction. It results in the transformation of the quasi-linear variation of $\Delta R$ with the external OOP magnetic field into a sharp switch as presented in Fig.~\ref{Fig3CoLayer} (b). Moreover, the switching accounts for about two third of the total amplitude of the $\Delta R$ signal, which indicates that the switching of the topmost Co magnetization in its gate-induced PMA state is coupled to the switching of magnetization in the middle Co layer. Subsequently, a positive gate voltage of $+$2.8\,V is applied to the stack during 5\,min that reverts the topmost Co magnetization back into the plane with an IP anisotropy state that is stronger compared to the as grown state, in agreement with the results presented in Fig.~\ref{Fig1CoLayer} (b). For both positive and negative gate voltages, the anisotropy of the two buried Co layers remains seemingly unchanged. The reversibility and cyclability of these effects are presented in Fig.~\ref{Fig1CoLayer} (c). The Hall resistance at remanence $R_\text{rem}$ is switched between two states with about 90\% and 66\% remanence after the application of subsequent positive and negative gate voltages, corresponding to the PMA and IP anisotropy states, respectively. Note that the first positive gate voltage is $+$2.8\,V, while for the following steps, a gate voltage of $+$3\,V is used. It should be noted that the remanence obtained in the PMA state is expected to be underestimated, since the unbiased IP anisotropy regions at the contacts also contribute to the total signal \textcolor{black}{(See the Hall resistance measurement configuration schematic in Supplementary Information \cite{SuppMat} Fig.~\textbf{S1})}. In Figure~\ref{Fig3CoLayer} (d), we show the same experiment measured using Kerr effect, in which the light is placed at the centre of the gating area, confirming the full remanence present in the PMA state. Moreover, it has to be pointed out that these Kerr effect measurements have performed several weeks after the gate voltage application and complete removal of the top ionic liquid and ITO gate, showing the reliability of the non-volatile effects both in the IP and PMA regimes. The same results are obtained for hysteresis loops measured by AGFM using an in-plane field (see Supplementary Information \cite{SuppMat} Fig.~\textbf{S5}).

The dynamics of the gate-induced spin reorientation transition between PMA and IP magnetic anisotropy in the topmost Co layer has been imaged in real time using Kerr microscopy. First, the top layer magnetic anisotropy is set to PMA using a negative gate voltage of $-$1.5\,V and the magnetization is saturated under a positive magnetic field. In this state, magnetic domains are nucleated in \textcolor{black}{the three Co layers of }the multilayer by applying a negative magnetic field. Then, a positive gate voltage of $+$2.5\,V is applied during a total gating time of 10\,min. The gate voltage is reduced to $+$2.5\,V from the $+$3\,V used in the previous experiments to induce a slower transition between PMA and IP anisotropy in the top Co layer. In Figures~\ref{Fig3CoDW} (a-b), we show Kerr microscopy images obtained though the transparent ionic liquid - ITO gate while applying the $+$2.5\,V gate voltage to the stack for different gating times: 5\,min (a), 7\,min (b) and 10\,min (c). It can be seen that a pronounced magnetic contrast between merged circular structures, appearing in dark grey, and the background, which correspond to the PMA domains nucleated using magnetic fields. A \textcolor{black}{subtler} contrast change is seen to propagate from right to left, making the swept areas appear in a lighter colour, regardless of the presence of PMA magnetic domains. \textcolor{black}{Four contrasts are observed in Figs.~\ref{Fig3CoDW} (a-c), indicating that the domains remain in the bottom Co layer independently of the IP or OOP magnetic anisotropy character of the topmost Co layer.  The corresponding magnetic configurations are schematized in panels (a) and (c).} This voltage gated-induced moving front is originated at the top right of the image and moves with a velocity of approximately 61\,$\upmu$m/min, as shown in the insets of Figs.~\ref{Fig3CoDW} (a-c). The intermediate state reached after the application of a $+$2.5\,V gate voltage for 10\,min (see Fig.~\ref{Fig3CoDW} (c)) is non-volatile. In Fig.~\ref{Fig3CoDW} (d),  we observe that a fully remanent hysteresis loop compatible with a large PMA is obtained by measuring only in the bottom left area of the Kerr image (blue curve), while the hysteresis loop of the entire image shows a significant decrease in the remanence (red curve), indicating that the top-right area of the image adds an IP anisotropy component to the total Kerr effect signal. The inset of Fig.~\ref{Fig3CoDW} (d) displays a Kerr image at remanance after the application of a positive external magnetic field showing distinct contrasts between the OOP bottom left area and the IP top-right area of the image. These measurements show that the gate-driven transition between PMA and IP anisotropy happens through the creation of a single 90$^{\circ}$ domain wall (DW$_{\updownarrow,\leftrightarrow}$) between the OOP and IP regions, and its propagation across the top Co layer. The motion of this DW$_{\updownarrow,\leftrightarrow}$ can be precisely controlled using the gate-voltage, and its position remains fixed once the gate-voltage is turned off. This opens the way to the design of spintronic devices exploiting all-electrical and ultra-low power control of the propagation of PMA-IP anisotropy domain walls. In particular, the electrical control of the nucleation and position of these domain walls where the DW chirality is fixed through the sign of the DMI could lead to new functionalities in magneto-logic devices \cite{Luo2020} and gives interesting perspectives for the study and optimisation of the dynamics of the magneto-ionic effects.

\begin{figure}[ht!]
\centering
\includegraphics[scale=0.7]{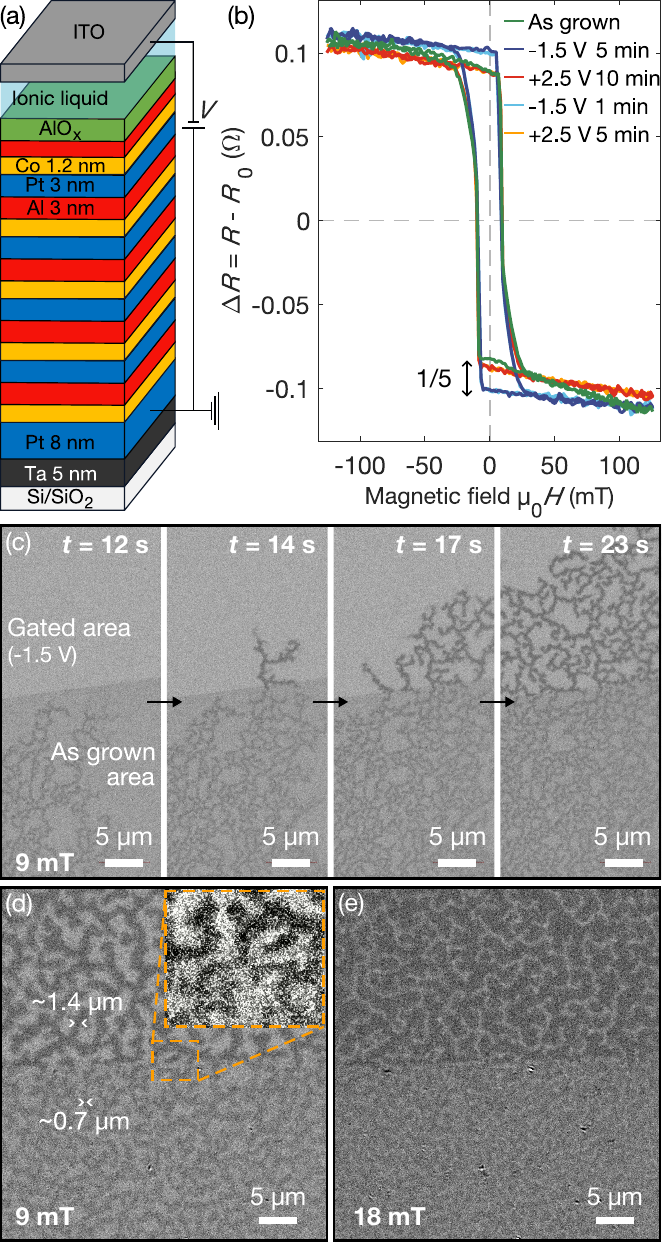}
\caption{(a) Magneto-ionic device composed of a Ta/Pt/[Co/Al/Pt]$_4$ /Co\textcolor{black}{/Al/AlO$_\text{x}$ ($y$ = 0.8\,nm)} multilayer stack. (b) Anomalous Hall resistance $\Delta R = R  - R_0$ measured after the application of several gate voltages. (c) Kerr microsopy images taken at the boundary between the as-grown (bottom) and gated (top) areas under an OOP magnetic field of 9\,mT showing the propagation of domains from the as-grown region into the gated region after 12\,s, 14\,s, 17\,s and 23\,s. (d-e) Kerr microsopy images of the final domain pattern at the boundary between the as-grown (bottom) and gated (top) areas obtained under an OOP magnetic field of 9\,mT (d) and 18\,mT (e). The inset in (d) shows the continuity of the domain pattern at the boundary. }
\label{Fig5CoLayer}
\end{figure}

To confirm that magneto-ionic gating can induce a variation in $m$ between $m = n$ and $m = n + 1$ in more extended multilayer structures, magneto-ionic devices have also been fabricated using multilayers with a number of repetitions of $n =$ 4, as shown in Figure~\ref{Fig5CoLayer} (a). In agreement with the results presented for $n =$ 2, the magnetization of the topmost Co layer undergoes a reversible and non-volatile spin reorientation transition as a function of the applied gate voltages. In Figure~\ref{Fig5CoLayer} (b), we display the anomalous Hall resistance $\Delta R$ hysteresis loops obtained in the as grown state, as well as for several subsequent applications of negative ($-$1.5\,V) and positive ($+$2.5\,V) gate voltages. We find that the remanence is modulated by 1/5 of the total $\Delta R$ signal, which is in agreement with the contribution from the topmost Co layer. It thus shows that magneto-ionic gating can change the number of repetitions of Co layers with PMA from $m =$ 4 to $m =$ 5. Interestingly, unlike the system with $n =$ 2, in this 5-repeats multilayer, only one magnetic switching is observed for both $m =$ 4 and $m =$ 5. This can be explained by an enhanced dipolar coupling between the four/five layers of Co that results in the collective switching of all coupled layers in a single event. \textcolor{black}{The cyclability of the voltage gate induced modification of the anisotropy in the topmost Co layer is reported in Supplementary Information \cite{SuppMat} Fig.~\textbf{S6}.}

In Figures~\ref{Fig5CoLayer} (d-e), we show Kerr images of the domain structure in the Ta/Pt/[Co/Al/Pt]$_4$/Co/\textcolor{black}{Al/AlO$_\text{x}$ ($y$ = 0.8\,nm)} stack under OOP magnetic fields of 9\,mT and 18\,mT, respectively. The images are taken at the edge of a gated area, after removal of the ionic liquid gate. The top area is exposed to a gate voltage of $-$1.5\,V for 1\,minute, inducing PMA in the topmost Co layer, while the bottom area of the image remained in the as-grown state (the hysteresis curves of both areas are provided in Supplementary Information \cite{SuppMat} Fig.~\textbf{S8} (a)). The gate-induced change in the number of Co layers contributing to the PMA from $m =$ 4 and $m =$ 5 has a significant impact on the magnetic domain structure. As shown in Fig.~\ref{Fig5CoLayer} (c), we see that the domains first nucleate in the as-grown region ($m =$ 4) under an external magnetic field of $\sim$9\,mT, and then propagate into the gated area ($m =$ 5) by crossing the sharp boundary between the two areas. 
As seen in Fig.~\ref{Fig5CoLayer} (d), the domain structure found in the gated $m =$ 5 region shows a maze structure with domain widths of about 1.4\,$\mu$m, while the width in the as-grown area with $m =$ 4 is reduced to about one half of that value. Despite the strong difference in domain pattern, a continuity between the domains in the two regions has been observed, as illustrated in the inset in Fig.~\ref{Fig5CoLayer} (d). For higher magnetic fields, the size of the maze domains decreases in both areas keeping the ratio and continuity, as depicted in Fig.~\ref{Fig5CoLayer} (e). This behavior has been found to be reproducible as shown in supplementary Information \cite{SuppMat} Fig.~\textbf{S8} (b). Such a strong variation in domain width can be understood in terms of the change in the number of PMA coupled layers from $m =$ 5 to $m =$ 4. This agrees with previous studies showing that the domain width in multilayers can rapidly \textcolor{black}{decrease} with the number of repeats, as a result of the interplay between domain wall and magneto-static energy \cite{Sbiaa2010, Fallarino2019}. Note that similar contribution of the PMA Co layers to the hysteresis loops have been observed between the stack with $n =$ 4 topped by a Co layer with IP magnetic anisotropy and a comparable stack with $n =$ 4 but without the top Co layer (see Supplementary \cite{SuppMat} Information Fig.~\textbf{S9}. The gate-driven control of $m$ presented here is therefore of high technological interest, since it uses the gate-controlled spin-reorientation transition in one layer of Co to bias the properties of an entire multilayer structure, introducing a dynamic control of properties linked to the number of repeats. 

In conclusion, we report on a pathway for gate voltage control of the magnetic anisotropy of a multilayer stack made of [Pt/Co/Al] repetitions. The magnetic anisotropy of Ta(8\,nm)/Pt(8\,nm)/Co(1.2\,nm)/Al($y$\,nm) with the top Al layer oxidized in air, is found to vary from PMA to IP magnetic anisotropy by changing the top Al layer thickness $y$ between 0.6\,nm to 1\,nm, which can be ascribed to the varying distance between the Al/AlO$_\text{x}$ front and the Co interface. Similar changes from PMA to IP magnetic anisotropy are achieved by applying positive and negative gate voltages to the stack with $y =$ 0.8\,nm, induced by the electric field-driven displacement of the oxidation front. We observe similar changes in magnetic anisotropy, transitioning from PMA to IP, in the topmost Co layer of Ta/Pt/[Co/Al/Pt]$_n$/Co/Al/AlO$_\text{x}$ stacks with ($n =$ 2, 4), showing the ability to modify the number of Co layers with perpendicular anisotropy $m$ between $n$ and $n +$ 1. Importantly, these modifications are found to be non-volatile. The changes in the remanent anomalous Hall effect total amplitude account respectively for 1/3 and 1/5 for the samples with $n =$ 2 and $n =$ 4. Moreover, controlling the anisotropy of the top Co layer translates into the control of the magnetic domain pattern of the coupled multilayer structure. This is evidenced in multilayers with $n =$ 4 in which a gate-induced increase to 5 Co layers with PMA produces a reduction by a 2-fold reduction in the domain periodicity within the maze domain structure. The domains are found to first nucleate in the area with smaller $m$, before propagating into the $m = n +$ 1 area through a sharp boundary between the two regions. These results offer promise for anisotropy and magnetic domains manipulation by gate voltage in devices made of multirepeat stacks, holding promise for voltage control of the skyrmion nucleation and motion in devices for memory, logic or neuromorphic applications.\\

\noindent
\textbf{Acknowledgements}
This work has been supported by the Horizon2020 Framework Program of the European Commission under FET-Proactive Grant SKYTOP (824123), by the European Research Council ERC under Grant bioSPINspired 682955, by the French National Research Agency (ANR) with grants TOPSKY (ANR-17-CE24-0025) and SplaSy (ANR-21-CE24-0008-01), by a France 2030 government grant (ANR-22-EXSP-0002 PEPR-SPIN-CHIREX) and by the JSPS KAKENHI Grant (21H05016).\\

%





\end{document}